# Blockchain-enabled tokenization for sustainable and inclusive infrastructure investment

**Yifeng Tian**[1], **Chaofeng Wang**[1], **Ashish Asutosh**[1], **Junghoon Woo**[1], **and Peter Adriaens**[2]

[1] College of Design, Construction, and Planning, University of Florida, Gainesville, FL 32603, USA
[2] Department of Civil and Environmental Engineering, University of Michigan, Ann Arbor, MI 48109, USA

Corresponding author: Yifeng Tian, yifeng.tian@ufl.edu +1(201)716-9955 School of Construction Management, University of Florida, Gainesville, FL 32603, USA

## Abstract

Infrastructure is critical for enabling society to function and the economy to thrive, but there is an increasing mismatch between the need for infrastructure investments and available capital, which is in consequence of constraints on public resources and limited capacity to leverage the private sector co-financing under the current system. With the emergence of distributed ledger technology, such as blockchain-enabled tokenization, there is a significant potential to improve investment liquidity, transparency, efficiency and create new economic models to integrate non-financial values to promote sustainability and inclusiveness. This research analyzed 21 projects to investigate how tokenization is implemented in energy infrastructure projects. Exploratory case study analyses were conducted, which shows the diversity of tokenization arrangements. The state of the art, potential benefits, implications, and obstacles associated with the application of tokenization in infrastructure investment and development are discussed. The purpose of this research is to understand tokenization within the context of the energy sector but also to forecast its application in a broad spectrum of infrastructure projects (e.g., transportation, telecommunication, healthcare, education).

Keywords

Blockchain-enabled tokenization; Infrastructure investment; Decentralized finance; Sustainability and inclusiveness; Fintech; Renewable energy transition

## 1. Introduction

Infrastructure is essential for social development and economic expansion. From power generation facilities to transport systems and water networks, infrastructure provides foundational services which enable society to function and the economy to thrive. Infrastructure plays a vital role in strengthening inclusiveness and sustainability (Gupta & Vegelin, 2016).

Investment in infrastructure is one of the main drivers of generating long-term growth and stimulating economies out of recession after a systemic crisis like the COVID-19 pandemic and contributes to the United Nations Sustainable Development Goals (SDGs) (Thacker et al., 2019).

Despite the fact that infrastructure is at the nexus of economic and social prosperity, there is an increasing mismatch between the need for more qualifying infrastructure and available financing. According to the World Economic Forum forecast, the world is facing a $15 trillion infrastructure gap by 2040 (Davisson & Losavio, 2020). The World Bank estimates that developing countries need to triple the current annual spending on infrastructure over the next decade (Mapila et al., 2017). The demand for investment in infrastructure will only increase along with time to meet the SDGs.

Under the current financing system, public sector finance through direct grants, subsidies, and concessionary loans is subject to public budget constraints and political influence (Gadenne, 2017). The budgetary constraints caused by the emergent pandemic spending reduction in long-term bank loans due to tighter financial regulations (e.g., Basel III) have dwindled the traditional public capital sources to fund infrastructure (Humphreys et al., 2018). While private investment in infrastructure, typically through project loans, bonds, and private equity, is only accessed for certain types of infrastructure and limited to a narrow set of institutional investors such as pension funds and insurance companies, thereby leaving a significant amount of capital and resources on the sidelines (Yescombe & Farquharson, 2018). The current infrastructure financing system also struggles with integrating environmental, social, and governance (ESG) factors or effectively offering incentives to promote SDGs (Thacker et al., 2019). Pioneering thinking and groundbreaking financing methods are required to ameliorate public sector engagement and mobilize private sources to bridge the widening infrastructure gap and realize inclusive and sustainable growth to positively impact society, economy, and environment.

The pace of innovations is accelerating exponentially in the information age. Leveraging new technologies, infrastructure financing and operating systems can be more intelligent, efficient, and resilient. Since the advent of Bitcoin in 2009, the profile of blockchain – a combination of distributed ledger technology (DLT) with a variety of block-based encryption technologies – has soared (Nakamoto, 2008). Blockchain is capable of maintaining permanent and tamper-proof records of data. With the advancements in the blockchain, a decentralized, immutable, and trusted system can be built, bringing improved security and transparency (Swan, 2017). Over the last decade, blockchain has gone from a promising concept to the technology billed for the future. In its evolution, blockchain technology has inspired the creation of new business models and caused major stirs in various industries (Morkunas et al., 2019; Woo et al., 2021). Building on the blockchain, tokenization enables the transition of assets with values in conventional forms or access rights into cryptographic tokens (Morrow & Zarrebini, 2019; Khan et al., 2020). The transition could improve efficiencies by orders of magnitude. Even though the application of tokenization in infrastructure development is scarce at present, it has shown great potential to serve as an alternative financing vehicle to supplement the current infrastructure development and finance system to bridge the gap (Tian et al., 2020).

Despite the growing attention from media sources and practitioners, potential applications of tokenization in infrastructure projects have not attracted comparable academic interest. It lacks a deeper grounding in theoretical and empirical analysis. This research explores the fundamentals of the emerging blockchain-enabled tokenization phenomenon in general. More specifically, it focuses on how best to employ this new technology in infrastructure investment and development, especially in the energy sector. The objective of this research is to understand how tokenization is currently applied in energy projects and to forecast its potential applications in a broader spectrum of infrastructure projects. In the following sections, relevant literature will be reviewed to discuss the critical factors in tokenization. The case study approach will be employed to analyze how tokenization is applied in 21 energy infrastructure projects. Subsequently, the research findings will be examined, and conclusions will be drawn.

## 2. The conceptual background of tokenization

Tokenization was used to describe a process when a sensitive data element is substituted with a non-sensitive equivalent, referred to as a token, with no extrinsic or exploitable meaning or value. In this case, tokenization is similar to encryption (Stapleton & Poore, 2011). With the advent of DLT, tokenization got a broader meaning, which describes the process of converting assets or access rights into cryptographic tokens on a blockchain (Nassr, 2020). In this case, tokenization is similar to securitization. Cryptographic tokens can represent assets as stores of value or access rights, such as shares in a company, ownership of a piece of real estate property, permissions to a platform, project bonds, or electricity produced by energy plants. In theory, any assets or rights can be tokenized and represented on a blockchain. Tokenization builds connections between the off-chain world and the on-chain world, where transaction efficiency, information recording, and sharing are expected to be improved (Laurent et al., 2018).

Cryptographic tokens are governed and executed through smart contracts, which are software algorithms integrated into a blockchain with trigger actions based on predefined parameters. Smart contracts are self-enforcing and self-executing (Zou et al., 2019). The automation reduces the administrative burden and the number of intermediaries required in the process, which leads to cost reduction and faster execution. Besides the efficiency gains brought by automation and disintermediation, tokenization delivers other benefits, including enhancing transparency, improving liquidity for currently illiquid assets (e.g., infrastructure assets), and enabling more efficient clearing and settlement (Wang et al., 2018). Cryptographic tokens can be classified as follows, payment tokens, utility tokens, and security tokens. Tokens can be issued through Initial Coin Offering (ICO), Security Token Offering (STO), IEO (Initial Exchange Offering), etc. (Myalo, 2019).

### 2.1. Token issuance

ICO

ICO is a type of capital-raising activity in the environment of the blockchain (Zetzsche et al., 2017). Cryptographic tokens are issued to raise capital for the creation of a blockchain or funding a blockchain-related venture (e.g., business or platform). With the successful issuance of the Ethereum blockchain in 2014, numerous ICOs were launched within a short period of time. Most ICOs were unregulated. Navigating securities laws have been critical, particularly for genuine utility virtual assets that are not intended to operate as investments but are more akin to a pre-paid digital coupon (Deloitte, 2020). Tokens issued through ICOs are often not registered as securities, although many have subsequently been determined to be investment products. Securities sold in ICO are illegal unless they meet regulatory requirements. Issuers have the flexibility to choose their ICO structure, including the price and quantity of tokens offered. Token holders do not necessarily have a claim to any assets of the company. Tokens' value is typically dependent upon the rights represented by the token and the development status of any underlying project. The issuer usually prepares a whitepaper detailing the business plan, financials, and other information about the project (Fenu et al., 2018). The ease with which money was raised also attracted market participants with bad motives who took advantage of the unregulated ICO market to scam investors (Momtaz et al., 2019). The value of tokens in early ICO projects is highly subject to speculation.

STO

STO is a type of public offering for tokenized digital securities, known as security tokens (Lambert et al., 2021). Similar to the concept of an Initial Public Offering (IPO), a company can raise funds by creating and issuing a security in the digital format to investors. In order to be sold as securities to investors, security tokens must meet the requirements under applicable securities regulations such as the Securities and Futures Ordinance in Hong Kong or the Securities Act of 1933 in the United States (Deloitte, 2020). Security tokens can be digital representations of any assets or instruments defined as security, such as stocks, bonds, real estate, intellectual property, etc. (Kranz et al., 2019). Issuers of security tokens must factor in relevant legal and regulatory requirements, as do brokers and exchanges. Security tokens may be listed on token exchanges or conventional stock exchanges and may be restricted only to professional or accredited investors. STO brings together the benefits of blockchain for financing but in a regulated environment, with the possibility of asset-backed structures increasing potential appeal. STOs became popular after the 2017 ICOs mania, which are associated with many scams and frauds (Momtaz et al., 2019). However, STOs didn't experience explosive growth like ICOs. Lacking regulatory clarification and supporting technological infrastructure of this emerging asset class impedes its adoption. Some security tokens are still issued through ICOs to avoid regulation.

IEO

IEO is offered by a token exchange on behalf of blockchain ventures to raise funds in token sales (Furnari, 2021). IEO is essentially ICO, but token trading platforms step in to fill the role of traditional securities distributors and conduct due diligence services on the crypto asset to ensure the credibility of the project. One of the biggest advantages of an IEO is token liquidity. Once the IEO ends, investors can start trading its tokens instantly on the exchange. The due diligence performed by the exchange is intended to provide credibility to the offering by mitigating risks (Deloitte, 2020). Nevertheless, not all jurisdictions regulate token exchanges. They may not be licensed by local regulators and/or the offering may be unregulated. In an unregulated regime, standards of diligence and disclosure may vary and investors' protection under the existing securities laws and regulations may be limited.

## 2.2 Public and private blockchain

A public blockchain network is permissionless, which allows all nodes of the blockchain to have equal rights to access, create, and validate new blocks of data (Guegan, 2017). The network typically has an incentivizing mechanism to encourage more participants to join the network. Public blockchains are decentralized and anonymous. However, they imply little to no privacy for transactions (Lai et al., 2018). A private blockchain network requires invitations and must be validated by either the network starter or by participants restricted by the starter. It is generally set up as a permissioned network (Pongnumkul et al., 2017). Once an entity has joined the network, it will play a role in maintaining the blockchain in a decentralized manner (Woo et al., 2020). On the downside, when it comes to security, private blockchains are more prone to security threats and other vulnerabilities because it has fewer nodes; thus, bad actors can gain access quickly (Hao et al., 2018). The majority of the current tokenization projects are based on Ethereum, Binance, or Solana blockchains, which are public blockchains. While a few private blockchain projects have also been launched successfully, such as World Bank's Bond-i.

## 2.3 Centralized and Decentralized token exchanges

Centralized token exchanges (CEXs) are trusted third parties to facilitate token trades relying on a private infrastructure to match supply and demand. Established reputations are leveraged to bring investor confidence (Lup et al., 2019). CEXs provide traditional forms of security, some in the form of insurance, others in the way of regulatory compliance, such as Know Your Customer (KYC) verification and anti-money laundering (AML) provisions. Fiat-to-crypto or crypto-to-crypto trading pairs are offered by CEXs. Binance, Coinbase, Kraken, and Huobi Global are top CEXs ranked by trading volume (Barbon & Ranaldo, 2021).

Decentralized exchanges (DEXs) are marketplaces (permissionless) that facilitate the peer-to-peer exchange of crypto, which do not involve a third party (Luo et al., 2020). DEXs use smart contracts and algorithms that self-execute once preset conditions are met. DEXs allow traders to execute trades with a certain degree of anonymity (i.e., pseudonymity) (Lin, 2019). The barriers to entry in DEXs are very low. Small-scale projects are easier to have their projects listed compared to CEXs, but a lack of screening process results in having lower quality projects listed. DEXs are in the early development stage. They have struggled for years to create levels of liquidity similar to centralized exchanges. Uniswap, Stellar, and PencakeSwap are leading DEXs in the market.

While DEXs represent a new technology that will likely become the standard in the future, CEXs are more user-friendly and theoretically safer at present (Victor & Weintraud, 2021). With the great influx of institutional investors into the crypto market, especially Bitcoin, it is safe to assume that centralized exchanges will stay ahead in the mid-term. However, leaps in technology, like decentralized finance (Defi), will keep pushing DEXs forward. Various hybrid models and complementary applications are likely to emerge (Chen & Bellavitis, 2020).

## 3. Case studies

This research adopts an exploratory case study approach to investigate how blockchain-enabled tokenization is applied in energy infrastructure investment and development. Given the fact that the body of tokenization knowledge is in its infancy, a case study approach is considered suited as it offers a detailed description of a phenomenon in particular instances, typically based on various sources of data. Theoretical propositions and constructs can be created by using case-based evidence (Yin, 1981; Yazan, 2015; Aberdeen, 2013). The multiple-case design enables a comprehensive examination and generalization of the results, which provides a holistic view of a phenomenon within its context (Stake, 2013; Maton & Salem, 1995). An exploratory case study aims at extending the understanding of phenomena in the early phase of development (Pan et al., 1999; Ogawa & Malen, 1991; Lam & Lam, 2016). It is applied when extensive empirical research has not been dedicated to the topic of interest. The use of this approach may be justified when the terrain is little known or stereotyped views are imposed. This approach is generally not rigidly structured but not based solely on assumptions. The purpose of the exploratory case study is not to generate definitive evidence but to suggest hypotheses to be investigated in future works (Flavell et al., 2019; Sali & Marasigan, 2020; Tian et al., 2022a).

Cases examined in this study were retrieved from desk research. Keywords used for searches of qualified cases are blockchain tokenization, asset tokenization, infrastructure investment, ICO, STO, energy, decentralized finance, renewable transition, transportation, and real asset. The criteria to define qualified cases are: (1) tokenization has to be the major technology applied in the project, (2) the project has to be in the infrastructure sector, and (3) tokenization plays a key role in facilitating the development of infrastructure projects or corporations in this sector. ICO/STO rating platforms (e.g., icobench, coindex, icodrops, storating, etc.) and relevant project news are the primary databases of this research. Thirty-five projects meet the requirement of the criteria. Fourteen projects were excluded due to missing essential information. Twenty-one tokenization cases in the context of the energy sector are identified. The use of various types of tokenization models in financing infrastructure corporations and projects across the globe is discussed. Tokenization projects are sampled and analyzed based on the types of tokens (payment, utility, and security). Potential and realized efficiency gains and obstacles faced in the adoption are discussed. Not only is this study conducted to understand the application of tokenization in energy projects but also to forecast its application in a broad spectrum of infrastructure projects (e.g., toll roads, airports, waste, telecommunication) and different financing mechanisms (e.g., corporate finance and project finance).

Case 1: SolarCoin

SolarCoin is a payment token designed to facilitate solar photovoltaic facility installation deployed on the SolarCoin blockchain (SolarCoin, 2021). It is an open community project run by the SolarCoin foundation, a Delaware-based US Public Benefit Corporation. SolarCoin was one of the earliest crypto-tokens offered to the public. Its ICO was in 2014. The foundation distributes SolarCoin to generators of solar electricity using verified solar facilities at the rate of one SolarCoin per one megawatt-hour (MWh) of solar energy produced. SolarCoin serves as a direct incentive to reward solar energy producers and encourages the development of solar infrastructure facilities and the transition to clean and renewable energy. SolarCoin can be used as a means of rewards and payment. SolarCoin has been losing popularity gradually. It was delisted from major exchanges in 2018. SolarCoin is now listed on secondary token exchanges (e.g., Carbonswap). After seven years of operation, SolarCoin was migrated to the Energy Web

Chain, an Ethereum-based chain, in April 2021. The estimated market capitalization of SolarCoin has decreased from its peak of around $100 million at the beginning of 2018 to less than $300,000 in December of 2021.

Case 2: WePower

Smart Energy Tokens are utility tokens issued by WePower, renewable energy procurement and trading platform founded in Lithuania, which aims at facilitating the global shift to renewable energy by democratizing the energy procurement process (WePower, 2021). Smart Energy Tokens represent fractional ownership rights and obligations obtained under the power purchasing agreement (PPA). WPR is another utility token issued by WePower. Both tokens are deployed on the Ethereum blockchain. WPR tokens allow access to WePower network functions such as donation pools and priority auctions. WPR tokens are listed on secondary token exchanges (e.g., Bitbns). The current WePower whitepaper was published in February 2019, the second version since the initial launch in 2018. WePower ICO raised $40 million in 2018. The estimated market capitalization of WPR is less than $2 million in December of 2021. WPR tokens have lost over 98% of their value at the time of launch.

Case 3. Cenfura

Cenfura Tokens (XCF) are utility tokens used as the medium of exchange and settlement for energy goods and services on the Cenfura Platform (Cenfura, 2021). Cenfura established the platform in 2020, an intelligent energy service company in the United Kingdom that develops and operates renewable energy assets. Energy purchasers and producers transact energy by using XCF through the platform. Cenfura claims to distribute its tokens to support inclusion and sustainability initiatives such as energy poverty programs. Cenfura targets the unbanked as a new group of stakeholders of its renewable energy projects. XCF tokens are listed on secondary token exchanges (e.g., Liquid). Daily trading volumes usually are less than $1,000 in December of 2021.

Case 4. ZiyenCoin

ZiyenCoins are equity-based security tokens backed by a fund of energy assets of Ziyen Energy, an American company founded in 2019 (Ziyen, 2020). The portfolio includes oil and gas reserves, minerals, and renewable energy assets. The rights to the company's Class A securities, incorporation certificates, and bylaws are embedded into each ZiyenCoin. ZiyenCoins were issued through security token offering (STO) and deployed on the Ethereum blockchain following ERC20 standards. The company was valued at $5 million at the time of token issuance. ZiyenCoins are not listed on any token exchanges. The price of each ZiyenCoin is at $0.01, arbitrarily determined by the management of the company.

Case 5. Electric Vehicle Zone

EVZ coins are utility tokens deployed on Ethereum as rewards or alternative means to pay for services operating on the EVZ platform (EVZ, 2021). The EVZ platform is an electric vehicle charge-sharing infrastructure that connects chargers in idle time and electric vehicle users, developed in South Korea in 2018. EVZ coins encourage the transformation of private-use charging points into public-use points and the transition to clean energy. EVZ coins are listed on secondary token exchanges (e.g., Bithumb). The price of EVZ coins has been gradually increasing over time since its issuance. The approximate return on investment if purchased at the launch time is about 800% in December of 2021.

Case 6. Green Life Energy

Green Life Energy tokens (GNL) are payment tokens deployed on the Binance Smart Chain issued by Green Life Energy Global, a British company founded in 2021 (GNL, 2021). The company's goal is to develop green energy projects globally. GNL token and wallet will be used in EV car charging port payment systems, purchasing power from renewable energy projects, investing in renewable power plants, and providing charitable support for other renewable projects. However, the company didn't reveal detailed plans for how GNL tokens will be applied in each scenario and how platforms or devices to be developed to facilitate the implementation of its token. GNL tokens are listed on a decentralized secondary token exchange, PancakeSwap. The daily trading volumes were around $10,000 in December of 2021.

Case 7. Impact PPA

Impact tokens (MPQ) are utility tokens deployed on the Ethereum blockchain issued by Impact PPA, an American company founded in 2017 (ImpactPPA, 2018). The Impact PPA platform allows energy consumers to "Pre-Pay" for electricity from a mobile device. MPQ tokens are sold to fund energy infrastructure projects. MPQ Token holders are granted access to review and vote on proposed projects embedded with the tokens. As claimed by the company, its platform and MPQ tokens could address issues of the unbanked, developing countries and facilitate the transition to renewable energy. However, the social media accounts of Impact PPA have remained inactive since 2018, and the proposed platform was not launched. MPQ tokens were not listed on any secondary token exchanges.

Case 8. SUNEX

SUNEX tokens are utility tokens deployed on the Ethereum blockchain issued by Sun Exchange, a South African company founded in 2015 (Sun Exchange, 2018). The SUNEX project was launched in 2018. Sun Exchange is a peer-to-peer solar leasing and crowdfunding platform. Investors worldwide can purchase solar photovoltaic (PV) cells and build wealth by leasing those cells to power businesses and organizations in emerging markets. SUNEX tokens are the native network token for the Sun Exchange platform. Funds raised from token sales are claimed to be used for pre-financing construction of solar projects, business development, and emerging market rural infrastructure project development. SUNEX ICO ended after $1 million was raised, which is 4% of the target. There have been few SUNEX updates since 2019. SUNEX tokens are not listed on any secondary token exchanges.

Case 9. NETZ

NETZ tokens are security tokens backed by renewable energy assets issued by Net Zero Enterprises, an American company founded in 1982 (NETZ, 2021). The NETZ project was launched in 2021. As claimed by the company, funds raised from NETZ sales will be invested in green energy power plants. Renewable energy assets will serve as collaterals of NETZ tokens. Revenue generated from the underlying plants will be distributed as dividends on a quarterly basis. However, there are no detailed plans of how investments in NETZ will be utilized and how financial information will be audited. NETZ STO was in 2021, 38 million tokens were sold, which is less than 50% of the target. NETZ tokens are listed on a decentralized secondary token exchange, Cointiger.

Case 10. SolarWind

SolarWind tokens (SLW) are utility tokens deployed on the Binance Smart Chain issued by SolarWind, an American company founded in 2021 (SolarWind, 2021). Funds raised from token sales will be used to finance solar projects. Project information will be available on a decentralized application (dAPP). The company claims to establish a nonprofit organization, Positive Energy, to support charitable missions voted by the SolarWind community. However, no information on the dAPP and the nonprofit organization can be found. Many milestones set by the company have not been achieved. SolarWind tokens are listed on two exchanges, PancakeSwap and Dex-Trade. The average daily trading volume in December of 2021 is less than $100.

Case 11. Solarex

Solarex tokens (SRX) are utility tokens deployed on the Ethereum blockchain issued by Solarex, a Britain company founded in 2017. The target markets of Solarex are in Africa, starting from Nigeria. Solarex tokens grant investors access to an ecosystem including Peer-to-Peer (P2P) energy trading, project financing, and other energy services. The company claims to allocate funds raised from token sales to support households without access to the national grid to finance solar projects in Africa. However, there is no information related to the recent development of the Solarex ecosystem that can be found. Solarex raised about $2 million in its ICO in 2019. SRX tokens were not listed on any secondary token exchanges.

Case 12. Restart Energy MWAT

MWAT tokens are utility tokens deployed on the Ethereum blockchain issued by Restart Energy, a Romanian company founded in 2015 (Restart Energy, 2021). The MWAT project was launched in 2017. MWAT Tokens holders are given access to the Restart Energy Democracy (RED) Platform to trade up to 1 MWh of energy per month. Tokenized energy traded on the RED Platform can be physically delivered at local rates in countries with deregulated energy markets, as claimed by the company. The platform is currently in operation. Funds raised from token sales are used to expand the platform. MWAT tokens are listed on a secondary token exchange, Kucoin. MWAT ICO raised $30 million in 2018. The price of one MWAT token was set as $0.1 at the time of token offering. It was traded at around $0.01 in December of 2021.

Case 13. Electrify.Asia

Electrify. Asia tokens (ELEC) are utility tokens deployed on the Ethereum blockchain issued by Electrify, a Singapore company established in 2017 (Electrify, 2021). ELEC Tokens holders are given access to Synergy's P2P electricity trading platform. ELEC tokens are used as transaction fees paid by energy providers and loyalty rewards for consumers. Synergy Alpha was tested in 2019. Synergy Beta was viable in 2021. Funds raised from token sales are used to support company development. ELEC tokens are listed on secondary token exchanges, Gate.io and HitBTC. ELEC ICO raised around $30 million. The price of one MWAT token was set as $0.08 at the time of token offering. It was traded at around $0.0001 in the December of 2021. It is worth mentioning that the price of one ELEC token went up from $0.0005 in January 2021 to 0.011 in April 2021, which was a 2000% increase in three months. The average market capitalization was less than $1 million in 2021.

Case 14. SunContract

SunContract Tokens (SNC) are utility tokens deployed on the Ethereum blockchain issued by Electrify, a Slovenian company founded in 2002 (SunContract, 2021). The SNC project was launched in 2017. SNC Tokens holders are given access to the P2P electricity trading platform, SunContract. The platform is currently in operation and has more than 5000 customers within Slovenia, as claimed by the company. ELEC tokens are listed on some secondary token exchanges, such as Huobi Global, HitBTC, and OKEx. SNC ICO raised around $2 million in 2017. The estimated market capitalization is around $3 million, and the average daily trading volume was about $200,000 in December 2021. The approximate return on investment in the December of 2021, if purchased at the time of launch, was about 30%.

Case 15. Megatech

Megatech tokens (MGT) are security tokens deployed on the Apollo blockchain issued by MGT Solar, a South African company founded in 2021 (Megatech, 2021). MGT Tokens are backed by equity shares of the company and the eight solar plants in development. However, the company doesn't reveal many details about the underlying assets and how those assets are linked with the tokens. Unaudited profit, 3000% Return on Investment (ROI), is posted on the website for marketing purposes. There is no information related to the projected solar plants that can be found. Funds raised from token sales will be allocated to develop solar power plants and the token system. The price of Megatech tokens is set at $0.01 at the time of offering. MGT tokens are listed on the secondary token exchange, BitMart. They were traded at $0.002 in the December of 2021. The price of Megatech tokens went up more than 3,000% overnight, traded at $0.045 on January 16, 2022.

Case 16. Enercom

Enercom tokens are security tokens deployed on the Ethereum blockchain issued by Enercom AG, a Liechtenstein company founded in 2016 (Enercom, 2021). The Enercom token project was launched in 2019. Enercom tokens are backed by equity shares of the company. Funds raised from STO, 200,000 tokens at the price of ETH 0.6 or EUR 100, are claimed to be used to acquire one telecommunication project and one renewable energy project. The whitepaper of the Enercom token is one of very few that explicitly explains risk factors, terms, and conditions of the securities and details of the offer/admission to trading. Token holders don't have voting rights but have rights to receive dividends. Enercom tokens are not listed on any secondary token exchanges.

Case 17. Sun Fund

Sun Fund tokens are security tokens deployed on the Ethereum blockchain issued by Sun Fund, an American company founded in 2017 (Sun Fund, 2019). Sun Fund develops, owns, and operates renewable energy assets. Sun Fund tokens are backed by revenue-generating renewable energy assets of the company. Investors participating in the token ICO also can receive preferred equity shares of the company and an additional simple agreement for future tokens (SAFT). The company claimed to build a producer-to-investor platform and Internet of Things (IoT) solar inverter to blockchain data channel oracle connectivity after ICO. However, Sun Fund stopped updating its social media profiles in 2018. No information on the company and the token project except ICO can be found. Sun Fund tokens are not listed on any secondary token exchange. Sun Fund ICO attracted 325 investors and raised $123,220.

Case 18. Grid+

GRID tokens are utility tokens deployed on the Ethereum blockchain issued by Grid+, an American company founded in 2017 (GridPlus, 2021). Grid+ is an eventual energy-sharing marketplace and conducting business in commercial electricity retail. Grid+ also manufactures hardware, programmable hardware wallet, as part of its business. Each GRID token is a credit on the Grid+ platform, redeemable by customers of the Grid+ platform for the right to purchase 500 kWh of electricity at the wholesale price available to Grid+. Grid+ operates with a two-token model. The BOLT token is treated by Grid+ as a stablecoin, redeemable by Grid+ customers for $1 worth of energy from Grid+ and backed by USD deposits. Grid+ raised $32 million in its ICO. GRID tokens are listed on secondary token exchanges, such as Balancer, Bancor Network, and Uniswap. The estimated market capitalization of WPP was over $60 million in December of 2021, increasing from $6 million earlier time of that year.

Case 19. WPP

WPP tokens are utility tokens deployed on the Ethereum blockchain issued by WPP ENERGY GmbH, a Swiss company founded in 2009 (WPP, 2021). The WPP project was launched in 2018 and revised in 2019. The company describes itself as a repository for green energy and environmental technologies. WPP tokens serve as payment tokens for the company's green hydrogen and biofuel futures contracts trading platform, HyFi. WPP tokens are designed to represent one kilogram of hydrogen gas and are accepted as a payment method to purchase convertible preferred equity shares of the company. Sun Fund tokens are listed on secondary token exchanges, such as BitMart and Digfinex. Around $60 million was raised in the ICO in 2018. The estimated market capitalization of WPP was around $20 million in December of 2021, while it surpassed $166 million in April of the same year.

Case 20. Efforce

EFFORCE tokens (WOZX) are utility tokens deployed on the Ethereum blockchain issued by EFFORCE, an Italian company founded in 2018 (Efforce, 2021). The WOZX project was launched in December 2020. WOZX tokens function as the medium through which energy savings created on the EFFORCE platform are tokenized for use by any participant. Like carbon credits, the tokenized energy credits on the platform can be freely traded. It is not clear whether the platform has been completed at the time of writing. The company claimed that the WOZX token might be equivalent to EFFORCE's stock shares in the future. WOZX tokens are listed on several secondary token exchanges, including a major exchange like Huobi Global. The estimated market capitalization of WOZX is around $40 million in December of 2021, while it surpassed $160 million in March of the same year. Four hundred fifty million WOZX tokens at the price of $0.1 were sold during a private sale in June 2019. WOZX tokens were traded at $0.27 at the time of writing in December of 2021.

Case 21. Powerledger

POWR tokens are utility tokens deployed on the Ethereum blockchain issued by Powerledger, an Australian company founded in 2016 (Powerledger, 2021). The POWR token project was launched in 2016. Powerledger platform allows P2P energy trading, enabling participants in local areas to sell and distribute solar power to their neighbors. The POWR token allows application hosts and participants to use the platform. The whitepaper of POWR tokens was revised three times in 2017, 2018, and 2019. The dual token system was developed, including Sparkz tokens, priced in local currency and POWR. POWR ICO raised $34 million in 2017. POWR tokens are listed on over 20 secondary token exchanges, including several major ones like Binance, Coinbase, and Bittrex. The estimated market capitalization of POWR reached $360 million in November of 2021. POWR has the largest market capitalization among all token projects analyzed in this study.

## 4. Cross-case analysis

Table 1 illustrates fundamental case information, including types of tokens, blockchain deployed, listing status, year of token offering, location of the project, price range, and market capitalization (cap) range. Information is extracted from crypto-information disclosure platforms, such as CoinMartcap and Binance, and websites of projects. Project information from January 2014 to December 2021 is analyzed. The main purpose of this research is to investigate how tokenization is applied in pioneering projects in the infrastructure sector. However, energy is the only sector applying tokenization at the current stage. This research analyzes the application of tokenization explored by energy projects. Research findings are disseminated to understand how tokenization could facilitate energy infrastructure development and how the specific experience learned from energy projects can be generalized to a broad spectrum of infrastructure projects.

Table 1. Summary of Cases

| Project | Tiker | Type of Token | Blockchain | Active (Y/N) | Token Exchange | Location | Year | Price Range | Market Cap Range |
|---|---|---|---|---|---|---|---|---|---|
| **Ziyen** | N/A | Security | Ethereum | N | N/A | USA | 2018 | N/A | N/A |
| **SolarCoin** | SLR | Payment | Energy Web | N | Carbonswap | USA | 2014 | $0.003 - $2.4 | $0.2M - $100M |
| **Wepower** | WPR | Ultility | Ethereum | Y | Bitbns | Lithuania | 2018 | $0.002 - 0.22 | $1.8 - $75 |
| **Cenfura** | XCF | Ultility | N/A | Y | Liquid | UK | 2020 | $0.018 - .018 | N/A |
| **Powerledger** | POWR | Ultility | Ethereum | Y | Many | Australia | 2018 | $0.03 - $1.8 | $15M - $650M |
| **Green Life Energy** | GNL | Ultilty | Binance | Y | Pancake | UK | 2021 | $0.01 - $0.065 | N/A |
| **ImpactPPA** | MPQ | Ultility | Ethereum | N | N/A | USA | 2018 | N/A | N/A |
| **Efforce** | WOZX | Ultility | Ethereum | Y | Many | Italy | 2020 | $0.2 - $3 | $30M - $160M |
| **WPP** | WPP | Ultility | Ethereum | Y | Digifinex/BitMart | Switzerland | 2019 | $0.0005 - $0.055 | $0.01M - $30M |
| **Sunex** | SUNEX | Ultility | Ethereum | N | N/A | South Africa | 2018 | N/A | N/A |
| **Sunfund** | N/A | Security | Ethereum | N | N/A | USA | 2018 | N/A | N/A |
| **Grid+** | GRID | Ultility | Ethereum | Y | Many | USA | 2017 | $0.03 - $3 | $1M - $120M |
| **Enercom** | N/A | Security | Polymath | N | N/A | Liechtenstein | 2019 | N/A | N/A |
| **Netz** | NETZ | Security | N/A | Y | Gate | Sweden | 2021 | N/A | N/A |
| **Megatech** | MGT | Security | Apollo | Y | BitMart | South Africa | 2020 | N/A | N/A |
| **Solarwind** | SLW | Ultility | Binance | Y | Dex-Trade | USA | 2021 | $0.001 - $0.025 | N/A |
| **Solarex** | SRX | Ultility | Ethereum | Y | N/A | UK | 2019 | N/A | N/A |
| **Electric Vehivle Zone** | EVZ | Ultility | Ethereum | Y | Bithumb | Korea | 2018 | $0.01 - $0.22 | $5M - $100M |
| **Restart Energy Mwat** | MWAT | Ultility | Ethereum | Y | N/A | Romania | 2017 | $0.001 - $0.05 | N/A |
| **Suncontract** | SNC | Ultility | Ethereum | Y | Multiple | Slovenia | 2017 | $0.01 - $0.65 | $1M - $70M |
| **Electrify.Asia** | ELEC | Ultility | Ethereum | Y | Gate/HitBTC | Singapore | 2017 | $0.0005 - $0.2 | $0.2 - $70M |

Among 21 projects included in this research, 20 of them were founded in developed countries defined by the UN (UN, 2014). Restart energy in Romania is the only project in developing countries. However, Romania is still considered a high-income country. Developing countries are lagging in this emerging field even though tokenization could potentially bring great benefits (Kshetri & Voas, 2018). There are some projects, like Solarex founded in the UK, that claim to target developing countries' markets. However, no business activities can be tracked. Tokenization is only making progress in developed countries at present. The SolarCoin ICO was in 2014, which was one of the earliest tokenization projects. Utility token ICOs were launched after 2017. Security token STOs were launched after 2018. This timeline reflects the general development process of tokenization, starting from cryptocurrencies (payment tokens) to utility and security tokens. The estimated market capitalization of these tokenization projects is relatively small, ranging from $100,000 to $300 million. The medium is less than $1 million at the token price in December of 2021. The tokenization market is rather volatile and the price of tokens can inflate a few times or depreciate by over 90% within a few days.

Cryptographic tokens are grouped into three categories depending on functions and purposes, payment, utility, and security (Wu et al., 2018). Solarcoin and Green Life Energy tokens are the only two payment tokens included in this study. Ziyencoin, Sun Fund, Enercom, Netz, and Megatech are the five security token projects. The rest of the 14 projects issue utility tokens. It is worth noting that some projects issue more than one type of token to facilitate transactions or simplify the management process (e.g., WePower, PowerLedger, Grid+, and EVZ). The categorization of the tokens is not always definitive. Some tokens issued by the project show characteristics of both payment and utility (e.g., green life energy token). In some cases, the purposes of the tokens evolve with the project. EFFORCE tokens (WOZX) are utility tokens as defined in its whitepaper, but they might be equivalent to the company's stock shares, which are securities, in the future.

### 4.1. Token models

Payment

Payment tokens are decentralized currencies for making and receiving payments on the blockchain, such as Bitcoin and Litecoin (Ferrari, 2020). Payment tokens are used as an alternative means of exchange and store of value, supplement to fiat currency. Payment tokens play an important role in facilitating transactions and reducing currency risks to attract investors. Payment tokens facilitate P2P transactions (Toderean et al., 2021). Token holders are able to send and receive payments to or from anyone on the network around the world at minimum costs and wait time. Due to token holders being able to send and receive tokens with only a smartphone or computer, The value of payment tokens is determined by the faith of its users in utility across the network (Nalder & Guo, 2020). Payment tokens gain or lose value on the basis of demand and supply in the market. The volatility of payment tokens is normally high. The price of one Solarcoin reached $2.7 in January of 2018, while it is trading at $ 0.008 in January of 2022.

Stablecoins are a class of payment tokens that attempt to offer price stability. Stablecoins have gained traction as they attempt to offer the best of both worlds: the instant processing and security or privacy of payments of cryptocurrencies and the volatility-free stable valuations of fiat currencies (Lipton et al., 2020). In the Grid+ case, BOLT tokens are issued as stablecoins, redeemable for $1 worth of energy from Grid+ and backed by USD deposits. It is difficult for payment tokens directly related to one specific sector to be successful due to the limited application scenarios to gain mainstream attention and sufficient users. Once Bitcoin has been accepted as the major medium of exchange in the world of crypto, which occupies the most share of the market. It is almost impossible for other tokens with a similar function to survive. Utility and security tokens can also be used as payment tokens, which further increases the difficulty of launching infrastructure-related payment token projects. SolarCoin has been delisted from most token exchanges, and it has lost over 99% of its value at the all-time high. Green Life Energy tokens are facing the same issues in gaining enough users (token holders) and offering valuable services to be alive and reach the goal of the developing team. Payment tokens issued within a platform at a smaller scale might be commonly used in future infrastructure tokenization projects like the Grid+ project.

Utility

Utility tokens grant holders access to use products or services offered by issuers of the tokens. They typically represent a contractual bundle of rights (e.g., the right to use a platform) stored on a blockchain initiative (Prat et al., 2019). The proceeds are commonly used to develop a platform and/or for other capital expenditure needs. Utility tokens are most often issued during a crowdsale or through ICOs (Crosser, 2018). Utility tokens are not security or currency. In most cases, utility tokens don't allow investors (holders) to attain power over a company's (issuer) decision-making process. The company's valuation is normally not associated with the value of the token (Pazos, 2018). Utility tokens are built around a community. The tokenization project backed by a solid community would contribute to the overall social and economic development from many perspectives and bring significant profits for issuers and investors. Due to the flexibility and efficiency gains offered by utility tokens, they have been experiencing exponential growth after 2015. Most tokens issued in ICOs are unregulated. The exemption from the existing security laws and regulations encourages innovations, but it also drastically increases the potential for scammers to exploit. Scam token issuers can take advantage of information asymmetry to profit through fraudulent means at minimum cost in an unregulated ICO.

The majority of the cases, 14 out of 21 (67%), analyzed in this research issued utility tokens. Those tokens grant holders access to a platform developed by the company, such as energy trading (e.g., Powerledger, WPP, WePower, Grid+), payment system (e.g., Electric Vehicle Zone), and crowdfunding (Sunnex). It is not necessary that platforms have been established before ICOs. Funds raised in ICOs are used to develop platforms or other supporting infrastructure. A successful ICO may just need a compiling story (reasonable business plan) and a trustworthy management team to convince investors to pitch in investments. It is very risky for investors to participate in utility token issuance when available information on the project is quite limited, which is very common. Some utility token projects succeeded, like Powerledger or Grid+, which bring investors over 1,000% investment returns if tokens were purchased at the time of issuance. However, for some projects (e.g., Impact PPA, Sunfund, Sunnex, or Enercom), projects failed before any platforms were built or any products or services were launched. If the tokens are backed by a failed project or community, investors could lose the entire investment.

Utility tokens are the most common among other types of tokens for companies issuing cryptographic tokens to raise funds. Lacking regulations could be one of the main contributors (Baranes et al., 2021). Without tedious regulations, project developers would face fewer hurdles to innovation. Most utility tokens issued in cases included in this research are unregulated, including those with the largest market capitalization (e.g., Powerledger). It certainly doesn't mean unregulated projects must be scams, but the risks that investors bear are high. Token issuers face minimum-to-zero consequences if their projects fail. The well-developed supporting infrastructure is another contributor to the fact that utility tokens are more popular compared to other types of tokens. Thousands of projects have issued utility tokens through ICOs since 2013. Resources around utility tokens issuance and management are ample. For token issuers, utility tokens are a more convenient and cost-effective choice at the current stage. Even though all tokenization projects studied in this research area are in the energy sector, tokenization might bring realizable efficiency gains in other sectors, such as transportation. Utility tokens can be backed by tolls to support the finance of projects.

Security

Security tokens are investment instruments representing assets of different classes, including equity, fixed income, real estate, structured product, investment fund shares, and commodities that are traded and held on a blockchain (a distributed ledger) (Mendelson, 2019). The Howey Test created by the U.S. Supreme Court, is a well-known test for determining whether an investment is a security and thus subject to securities regulations. The Howey Test can be applied to differentiate security tokens from payment and utility tokens. For a traditional asset that doesn't exist on a blockchain, security tokens can be created through asset tokenization. For assets that already have an on-chain presence, tokens can be created through asset origination. These assets are referred to as natively digital securities. They can be created by mining or staking, depending on the blockchain (Subramanian, 2019). The value of security tokens derives from the underlying asset. Security tokens are issued through a STO instead of an ICO. Issuance and transactions of security tokens are subject to security regulations and legislation (Rivero, 2018). Issuers of tokens must factor in relevant legal and regulatory requirements, as do brokers and exchanges. Security tokens may be listed on an exchange and may be restricted only to professional or accredited investors. The additional regulatory oversight makes them less vulnerable to fraud and misuse. The blockchain environment enhances securities regulatory objectives of disclosure, fairness, and market integrity and supports innovation and efficiency through automation enabled by smart contracts (Hines, 2020).

Security tokens are not without risk, as with any investment product or structure. Hype cycles certainly continue. Although the identity of the investor might be recorded in the smart contract, the position on whether virtual assets constitute "property", the definition of ownership, and requirements for the assignment of the virtual assets may lack legal clarity or involve overlapping and even conflicting laws at present (Nassr, 2020). The valuation of virtual assets remains difficult. Without historical performance data, accounting professionals have difficulty ascertaining the fair value of the security token (Lambert et al., 2021). And due to the unique design and structure of each security token, it is difficult to find a referencing token for valuation. Regulatory compliance is a double-edged sword. With greater regulatory protection and remedies, the risk of fraud is reduced (Liu & Wang, 2019). However, regulatory hurdles are one of the reasons that security tokens' growth and adoption have been rather modest in comparison with utility tokens' exponential growth. It is still not common for major securities exchanges to launch an STO or have security tokens listed.

Five out of 21 cases analyzed in this research issued security tokens. Four out of the five projects' security tokens are backed by the equity shares of the issuing company. Assets are held through a special purpose vehicle (SPV). Netz is the only project claiming that its tokens are directly backed by green energy assets. In theory, the value of tokens is determined by the value of the underlying company or assets. In the case of ZiyenCoin, the company is valued at $5 million. Five hundred million ZiyenCoins were issued at the price of $0.01. However, how the company or assets are valued remains unclear. None of the security tokens are listed on secondary token exchanges except the Megatech token. Security tokens are normally sold in private sales. Megatech claims that its tokens are equity tokens backed by the actual company, which should be categorized as security tokens. However, in its whitepaper, Megatech tokens are defined as utility tokens that are issued through ICO. Security regulatory compliance is not mentioned in its whitepaper. Some companies are still in the very early phase of exploring tokenization. There has been no uniform definition and classification in this field. Many security token projects claimed to be regulated. However, there is no evidence these projects are filed and being audited properly. Regulatory compliance is more like a gimmick to attract investors sometimes. Most whitepapers emphasize the estimated returns that investors could receive by investing in these projects. However, the information on the valuation of the company or assets, the business plan of the company, and the auditing and regulatory compliance process are limited. At the time of writing, none of the projects included in this research revealed information on how milestones set in the whitepaper has been reached. In other words, these companies have ceased to operate. According to data of blockchain analytics, very few transactions of ZiyenCoins were made after the mid of 2019. Information related to Sunfund and Enercom besides their whitepapers is scarce. Netz, as a new project founded in 2021, is struggling to reach its STO goals. Megatech is the only project that remains active, and its tokens are actively traded on a token exchange. The tokenization models applied in these projects are all corporate finance.

## 5. Discussion

### 5.1 Benefits

Transparency

The hallmark of blockchain technology is transparency. Information of transactions, addresses, and quantities of tokens are immutably stored on distributed ledgers. If it is a public blockchain, all internet users have real-time access to the information (Rizal Batubara et al., 2019). Better transparency is mentioned in all whitepapers of the tokenization projects studied in this research. Etherscan is a block explorer and analytics platform which allows users to search the Ethereum blockchain for activities taking place on chains. Other blockchains have their explorer platforms, like Binancescanner and Solscan. Real-time access to immutable data increases the confidence of users and simplifies the due diligence process. Suppose integrated with IoT devices, such as smart sensors installed on infrastructure facilities, financial or operational data can be automatically recorded and stored on a blockchain through an oracle (Reyna et al.,

2018). Users, such as governments, project sponsors, project administrators, financial practitioners, investors, and surrounding communities, can monitor the financial and performance information (if possible) of the project through data stored on the public blockchain. Some companies have explored integration. Sunfund claimed to build a producer-to-investor platform and IoT solar inverter to blockchain data channel oracle connectivity. The unprecedented scope and granularity of data streams could render the financial modeling of the project more accurate in predicting asset value. Timely and accurate data supports better regulatory compliance and decision-making. Even though on-chain data can be recorded immutably and available to be revealed in real-time, not all project information is available on the blockchain. The transparency brought by blockchain in these cases is mainly limited to data generated through on-chain activities, but off-chain information remains private.

Efficiency

Smart contracts allow for bi-directional instant transfer of funds and tokens by removing intermediaries without the need for a separate settlement process. The disintermediation enabled by smart contracts results in a significant fixed financing cost reduction and improved bankability of projects (Schar, 2021). The scale of projects or corporations is no longer a determining factor for financing cost efficiency. Small-medium enterprises (SMEs) valued at less than $10 million, such as Ziyen, Solarex, and Cenfura, are able to raise funds from the public directly to support their business activities and finance future infrastructure projects. It is not cost-effective for projects or corporations valued at less than $50 million to go public in the conventional financing system. Except for Powerledger and Grid+, all other projects issuing tokens to raise funds are under $50 million. Whether tokenization could reduce the cost of capital for projects over $100 million and bring enough efficiency gains to institutional investors needs further investigation. Smart contracts enhance governance by encoding contractual agreements and predefined conditions into programs. Unless conditions were met, the smart contract would not be executed, eliminating risks of breach of contract and consequences of legal actions. In theory, tokenization could enable the democratization of finance and make investments in small-scale infrastructure at the project level possible, greatly expanding economic opportunities and market growth of infrastructure.

Liquidity

Tokenization could improve the liquidity of investments in infrastructure projects by unlocking global markets to capture capital and facilitating trading in secondary markets. By allowing for the atomicity of transactions via hash time-locked contracts (HTLC), tokenization solutions remove geographical restrictions, which allows global participants, both project developers and investors, to be easily involved and make transactions anywhere and anytime (Tian et al., 2020b). Twenty-one projects in fourteen countries studied in this research are all available to international investors. Investors can invest in any projects meeting their investment goals without being restricted by their nationalities. Cryptographic tokens listed on secondary markets are available to trade twenty-four hours a day and seven days a week. Fifteen out of twenty-one projects have their tokens listed on token exchanges, including centralized exchange (e.g., Binance, BitMart) or decentralized exchange (e.g., Uniswap, PancakeSwap). Powerledger tokens are listed on more than twenty exchanges worldwide. The average daily trading volume in 2021 is over $10 million. It topped at $1.4 billion on November 22, 2021. While the average daily trading volume of other tokens with a smaller market capitalization, like Solarwind, can be as low as a few hundred dollars. The liquidity of these tokens is still limited even though secondary trading is enabled. Secondary trading also gives rise to derivatives building on top of tokens, which makes other value-added services to be easily realized. Options and futures trading is offered by some exchanges, like Binance and Huobi Global. Regulations of security tokens might undermine the liquidity opportunities as fewer investors are qualified to participate. In the case of ZiyenCoin, international investors can invest in these tokens pursuant to Regulation S. However, in terms of domestic investors, they have to be accredited to invest pursuant to SEC Rule 506(c) of Regulation D.

Sustainability and inclusiveness

By tokenizing security or utility into digital tokens, which represent a small portion of economic benefits or access rights, investing in infrastructure projects with a small amount of money becomes financially viable (Giungato et al., 2017). The lowered barriers to entry would attract small investors, including retail and SME investors, who were excluded from the infrastructure investment under the conventional financing system. The vast majority of investors participating in ICOs and STOs of the projects analyzed in this research are small investors. Institutional investors merely started to invest in crypto assets in 2021. By enabling investing with a small amount of money, surrounding community residents of the project might be offered the opportunity to be actively engaged in project financing. They can serve as special shareholders to influence the planning and future operations of the project through voting rights (designed by regulators and project developers) embedded with the tokens. As claimed by the company, the Slovenian SunContract platform has more than 5,000 registered customers within Slovenia holding SNC tokens. Local customers can easily trade energy on the platform. Consumerization through tokenization thus has the capacity to create a sense of local ownership in the services of the public infrastructure facilities and provides a platform to galvanize social acceptance. Even though this theory has not been proved in the real world since all projects studied in this research are in private hands, it still might be a viable option that could be considered by public projects in the future. Blockchain-based tokenization offers a potential solution to provide cost-efficient digital identity, credit history, and ownership recognition. With blockchain-run digital wallets and mobile payments based on smartphones, the blockchain solution can help onboard the unbanked to boost financial inclusion and growth. ImpactPPA and Cenfura have mentioned that the unbanked are targeted as a new group of stakeholders of their renewable energy projects. However, how it can be realized and whether the benefits brought by involving the unbanked can offset the extra efforts taken on them remain questionable.

Promoting sustainability and transition to renewable energy are the main targets of the projects studied in this research. These goals have been mentioned by the majority of them. Tokenization could create a new economic model and market to enable the conversion of non-financial values such as positive social and environmental impacts into security or utility tokens, which can be monetized and traded. Impact tokens are blockchain tokens backed by measurable social and environmental impacts (Uzsoki & Guerdat, 2019). They help stimulate the consumption of sustainable services and goods by providing financial incentives to users and customers, which improve the profitability and bankability of high-impact projects. However, what kinds of impacts are best suited to be converted into tangible economic benefits is still under exploration. In the SolarCoin case, the foundation distributes SolarCoin to generators of solar electricity using verified solar facilities at the rate of one SolarCoin per one megawatt-hour (MWh) of solar energy produced. SolarCoin serves as a direct incentive to reward solar energy producers. The economic benefits encourage the development of solar infrastructure facilities and facilitate the transition to clean and renewable energy. However, the SolarCoin project has been inactive since 2020. Investors didn't get on board and accept the vision of the project developers.

## 5.2. Challenges

Regulation

Most tokenization projects studied in this research are unregulated except for some security token projects which claim to be compliant with existing security laws. A potentially unclear regulatory and legal status for tokenization is a risk to all market participants, especially retail investors. The issue must be addressed by clarity and interpretation of existing or new regulations by financial regulatory and supervisory authorities. The unregulated environment explains the great number of scams and frauds in the ICOs. The tokenized market needs to comply with the regulations to protect investors and promote stability while facilitating healthy competition (Savelyev, 2018). It is not expected to raise significant issues in most jurisdictions since tokenization merely replaces the conventional electronic book-entries system with DLT-enabled networks (Nassr, 2020). Nevertheless, it is still uncertain which regulatory asset class and reporting structure tokenized assets are classified into. There are gaps between novel asset classes and business models created by tokenization and the existing regulatory framework. The lack of regulatory clarity for tokenized assets may result in arbitrage opportunities and an unsecured investment environment with a negative impact on all participants, thus delaying broader adoption of the tokenized markets (Schwerin, 2018). Even though the cross-border transfer is touted as a significant tokenization benefit for infrastructure investors, the alignment of international and domestic regulations is creating an implementation barrier. The legal status of smart contracts remains to be defined, as these are still not considered to be legal contracts in most jurisdictions. Although the identity of the investor might be recorded in the smart contract, the position on whether virtual assets constitute "property", the definition of ownership, and requirements for the assignment of the virtual assets lack legal clarity or involve overlapping and even conflicting laws.

Business models

Unrealistic expectations (e.g., financial projections and business plans) are frequently found in whitepapers of some projects studied in this research. These expectations built by some project developers may incentivize them to transition to DLT-based solutions without a proven rationale Mokunas et al., 2019). Unlike IPOs, companies have had a working product that's generating revenue before going public. ICOs/STOs only need to publish unproven concepts/ideas in whitepapers at the very early stage of project development. Project developers can easily take advantage of the "flexibility" to ignore reality and exaggerate outcomes to raise funds quickly from investors, especially when the market is highly speculative. However, once investors realize a project is unlikely to process, investments are drawn quickly. Liquidity can dry up in a few days. For many projects, there is not a solid business rationale for the application of tokenization (e.g., business problems are not clearly defined; major technical challenges are ignored; efficiency gains are insufficient to support the transition). Successful projects are more likely based on proven business models and moving small steps forward by integrating the benefits brought by tokenization with established systems. Revolutionary actions with the target of solving multiple and macro-level problems without detailed plans are likely to fail. Revisions in whitepapers are very common, which are made to address the technological breakthrough and changes in the market. The whitepaper of POWR tokens has been revised three times in 2017, 2018, and 2019. However, frequent major changes in the business plan may confuse investors and make them lose confidence in the project, although the execution is easy.

Technology

The adoption of tokenization at a large scale would face a number of challenges related to the underlying technology itself. Scalability is still a technological challenge of DLT-enabled networks and is relevant to tokenization given the significant throughput that would be required for the scale of global financial markets (Collomb & Sok, 2016). Interoperability between different networks needs to be secured for connectivity between markets to be allowed. Other technological risks include network stability and exposure to cyber risks but also business risks related to the migration to a DLT-enabled environment. Connections between on-chain tokens and off-chain underlying infrastructure assets are crucial in tokenization. A trusted central authority is required to guarantee the connection between the token and the asset, monitor the underlying infrastructure assets, and update the status if conditions are changed (Nassr, 2020). However, a system to guarantee the connection between the off-chain assets and on-chain tokens has yet to be established. Without the guaranteed connection, it is questionable whether the tokens are backed by the underlying infrastructure assets, which would significantly disrupt the credibility of tokenization markets and the private sector's participation.

### 5.3. Future applications of tokenization in infrastructure

Twenty-one tokenization projects are all in the energy sector. The applications in other sectors of infrastructure have not been found yet. Distributed renewable energy sources being naturally matched with the tokenization models explains why the energy sector is leading in applying this new financing method. Furthermore, most of the energy infrastructure is owned by the private sector, which indicates that the adoption of tokenization in the public sector (e.g., public infrastructure) is lagging. Private entities are normally acting faster and have more economic incentives to innovate. However, tokenization could be applied in other sectors. Utility tokens backed by tolls can be issued to raise funds for transportation projects. Bond-like tokens (security tokens backed by debts) can be issued to raise funds for public infrastructure projects. Depending on the project, whether it generates revenues, bond-like tokens can be paid back by project revenues or availability payments backed by the government in Public-Private Partnership (PPP) projects (Tian et al., 2020b). The financing mechanism of all the tokenization projects studied in this research can be categorized as corporate finance. Capitals raised from token sales are used to support corporate-level activities. In theory, it would face no issues with applying tokenization in project finance. An infrastructure project can issue tokens, backed project equity (in PPP projects), and project debts, to fund specifically this project. There have been no projects integrating tokenization with project finance at present, but it is a rather interesting new market to explore.

The application of tokenization can be referred to as the process of tokenizing utilities (e.g., services or products provided by the facility) or ownership/economic interests of the securities (e.g., equity of infrastructure companies or funds, loans, and bonds) on the blockchain. Given the increasing transition of infrastructure to intelligent and inclusive systems and the desire to unlock efficient financing, tokenization may facilitate establishing alternative financing models to overcome some obstacles that can't be addressed by conventional financing instruments. Table 2 illustrates the comparison of tokenization with conventional infrastructure financing instruments. Evaluations are based on the authors' estimation and analysis. Tokenization shows the potential to improve investment liquidity, transparency, and private participation. However, due to the absence of regulation, tokenized investment remains risky.

Table 2. Conventional Infrastructure Financing Instruments and Tokenization

| Host's View/Features | Direct government spending | Municipal and sub-sovereign bonds | Commercial loan | Listed equity/funds | Unlisted direct equity investment and co-investment platforms | Tokenization |
|---|---|---|---|---|---|---|
| Pros | No payback obligation | Low borrowing costs<br>High credit quality<br>Tax-free | Reliable funding source<br>Most applied | Direct access to the capital market | Direct ownership and management<br>Higher return | Expanded investor pool<br>Lower barriers to entry |
| Cons | Subject to political uncertainty<br>Limited availability | Low return rate<br>Default risks<br>Country risks | Highly Fragmented<br>Multiple intermediaries<br>High costs | Barriers to entry<br>High upfront and fixed fees | Limited liquidity<br>High-expertise required<br>High upfront investment | Regulation uncertainty<br>High risks and volatilities |
| Investment Liquidity | * | *** | * | *** | * | *** |
| Transaction Efficiency | ** | * | ** | * | ** | ** |
| Transparency | * | ** | * | *** | * | *** |
| Private participation | * | *** | * | *** | * | *** |
| Security | *** | *** | ** | ** | ** | * |

Note: ***indicates high applicability; ** indicates medium applicability; * indicates low applicability.

# 6. Conclusions

This study analyzed 21 projects to investigate how blockchain-enabled tokenization is currently implemented in energy infrastructure projects. Through the exploratory multiple case study analysis, this research shows the diversity of tokenization arrangements (payment, utility, and security). The state of the art, potential benefits, implications, and obstacles associated with the application of tokenization in infrastructure investment and development were discussed. The purpose of this research is not only to understand tokenization within the context of the energy sector but also to forecast its application in a broad spectrum of infrastructure projects (e.g., transportation, telecommunication, healthcare, education). It was found that tokenization showed the potential to improve investment liquidity, transparency, and efficiency and enable new business models to promote sustainability and inclusiveness. Nevertheless, tokenization is in its very easy stage of development. Challenges of regulations (e.g., unclear legal statutes, cross-border uncertainties, and lack of single point of accountability) and technology (e.g., network stability, interoperability, and cyber-risks) are preventing tokenization from realizing large-scale adoption.

To facilitate the adoption of tokenization to reach its full potential, policymakers need to update the legal system and regulation to address the changes brought by the emerging technology in a transparent and secure environment. Collaboration and education are considered critical steppingstones. Collaborations of public and private entities are critical. An open-access toolbox complied with case studies, lessons learned, and other educational materials should be established. Public infrastructure projects should explore applying tokenization to engage the surrounding community. At the same time, private entities should be encouraged to investigate how tokenization could be applied in infrastructure projects other than energy under PPP. Investments in Financial education for retail investors are necessary since they are included and could play an important role in the tokenized market.

Considering tokenization is in its infancy, there is limited data available for in-depth analysis at present. Future research should conduct analysis based on data from operations to explicitly examine to what extent and how tokenization can realize the projected efficiencies gains (comparing tokenized finance with conventional finance options). The application of tokenization in infrastructure projects other than energy should also be studied. The research on how tokenization can be integrated with project finance would contribute to the infrastructure finance literature significantly.

Tokenization can play a critical role in revolutionizing infrastructure finance and facilitating infrastructure investment to unlock efficiency gains in both the public and private sectors. Once the potential risks and challenges are carefully examined and mitigated, tokenization might promote economic growth, advance social equity, contribute to the climate agenda, and eventually achieve the SDGs in the long term. A coordinated, proactive, and long-term approach that builds on existing practices by the public and private sectors is required to maximize the potential benefits brought by the emerging blockchain-enabled technology.

# 7. References

Gupta, J., & Vegelin, C. (2016). Sustainable development goals and inclusive development. *International environmental agreements: politics, law and economics*, *16*(3), 433-448.

Thacker, S., Adshead, D., Fay, M., Hallegatte, S., Harvey, M., Meller, H., ... & Hall, J. W. (2019). Infrastructure for sustainable development. *Nature Sustainability*, *2*(4), 324-331.

Davisson,K. & Losavio, J. (2020). How sustainable infrastructure can aid the post-covid recovery. https://www.weforum.org/agenda/2020/04/coronavirus-covid-19-sustainable-infrastructure-investments-aid-recovery/#:~:text=The%20world%20is%20on%2Dtrend,by%201.5%25%20four%20years%20later.

Mapila, K., Lauridsen, M. & Chastenay, C. (2017). Mobilizing Institutional Investments into Emerging Market Infrastructure. *International Finance Corporation*. https://www.ifc.org/wps/wcm/connect/25529a56-643a-4c1a-adcd-21d9e9c48feb/EMCompass+Note+36+MCPP+FINAL+329.pdf?MOD=AJPERES&CVID=lJDT0aX.

Gadenne, L. (2017). Tax me, but spend wisely? Sources of public finance and government accountability. *American Economic Journal: Applied Economics*, 274-314.

Humphreys, E., van der Kerk, A., & Fonseca, C. (2018). Public finance for water infrastructure development and its practical challenges for small towns. *Water Policy*, *20*(S1), 100-111.

Yescombe, E. R., & Farquharson, E. (2018). *Public-private partnerships for infrastructure: Principles of policy and finance*. Butterworth-Heinemann.

Thacker, S., Adshead, D., Fay, M., Hallegatte, S., Harvey, M., Meller, H., ... & Hall, J. W. (2019). Infrastructure for sustainable development. *Nature Sustainability*, *2*(4), 324-331.

Nakamoto, S. (2008). Bitcoin: A peer-to-peer electronic cash system. *Decentralized Business Review*, 21260.

Swan, M. (2017). Anticipating the economic benefits of blockchain. *Technology innovation management review*, *7*(10), 6-13.

Morkunas, V. J., Paschen, J., & Boon, E. (2019). How blockchain technologies impact your business model. *Business Horizons*, *62*(3), 295-306.

Morrow, M. J., & Zarrebini, M. (2019). Blockchain and the tokenization of the individual: societal implications. *Future Internet*, *11*(10), 220.

Khan, N., Kchouri, B., Yatoo, N. A., Kräussl, Z., Patel, A., & State, R. (2020). Tokenization of sukuk: Ethereum case study. *Global Finance Journal*, 100539.

Tian, Y., Lu, Z., Adriaens, P., Minchin, R. E., Caithness, A., & Woo, J. (2020a). Finance infrastructure through blockchain-based tokenization. *Frontiers of Engineering Management*, *7*(4), 485-499.

Stapleton, J., & Poore, R. S. (2011). Tokenization and other methods of security for cardholder data. *Information Security Journal: A Global Perspective*, *20*(2), 91-99.

Nassr, IK. (2020). *The tokenisation of assets and potential implications for financial markets*. https://www.oecd.org/finance/The-Tokenisation-of-Assets-and-Potential-Implications-for-Financial-Markets.htm.

Laurent, P., Chollet, T., Burke, M., & Seers, T. (2018). The tokenization of assets is disrupting the financial industry. Are you ready. *Inside magazine*, *19*, 62-67.

Zou, W., Lo, D., Kochhar, P. S., Le, X. B. D., Xia, X., Feng, Y., Chen, Z., & Xu, B. (2019). Smart contract development: Challenges and opportunities. *IEEE Transactions on Software Engineering*, 47(10), 2084-2106.

Wang, S., Yuan, Y., Wang, X., Li, J., Qin, R., & Wang, F. Y. (2018). An overview of smart contract: architecture, applications, and future trends. In *2018 IEEE Intelligent Vehicles Symposium (IV)* (pp. 108-113). IEEE.

Myalo, A. S. (2019). Comparative analysis of ICO, DAOICO, IEO and STO. Case study. *Финансы: теория и практика*, *23*(6), 6-25.

Zetzsche, D. A., Buckley, R. P., Arner, D. W., & Föhr, L. (2017). The ICO Gold Rush: It's a scam, it's a bubble, it's a super challenge for regulators. *University of Luxembourg Law Working Paper*, (11), 17-83.

Deloitte. (2020). *Security token offerings: The next phase of financial market evolution?* https://www2.deloitte.com/content/dam/Deloitte/cn/Documents/audit/deloitte-cn-audit-security-token-offering-en-201009.pdf

Fenu, G., Marchesi, L., Marchesi, M., & Tonelli, R. (2018, March). The ICO phenomenon and its relationships with ethereum smart contract environment. In 2018 International Workshop on Blockchain Oriented Software Engineering (IWBOSE) (pp. 26-32). IEEE.

Momtaz, P. P., Rennertseder, K., & Schroeder, H. (2019). Token offerings: a revolution in corporate finance? *Available at SSRN 3346964*.

Lambert, T., Liebau, D., & Roosenboom, P. (2021). Security token offerings. *Small Business Economics*, 1-27.

Kranz, J., Nagel, E., & Yoo, Y. (2019). Blockchain token sale. *Business & Information Systems Engineering*, *61*(6), 745-753.

Furnari, S. L. (2021). Trough equity crowdfunding evolution and involution: initial coin offering and initial exchange offering. *Lex Russica*, (1 (170)), 101-117.

Guegan, D. (2017). *Public blockchain versus private blokhain.* https://halshs.archives-ouvertes.fr/halshs-01524440/document.

Lai, R., & Chuen, D. L. K. (2018). Blockchain–from public to private. In *Handbook of Blockchain, Digital Finance, and Inclusion, Volume 2* (pp. 145-177). Academic Press.

Pongnumkul, S., Siripanpornchana, C., & Thajchayapong, S. (2017). Performance analysis of private blockchain platforms in varying workloads. In *2017 26th International Conference on Computer Communication and Networks (ICCCN)* (pp. 1-6). IEEE.

Woo, J., Kibert, C. J., Newman, R., Kachi, A. S. K., Fatima, R., & Tian, Y. (2020). A New Blockchain Digital MRV (Measurement, Reporting, and Verification) Architecture for Existing Building Energy Performance. In *2020 2nd Conference on Blockchain Research & Applications for Innovative Networks and Services (BRAINS)* (pp. 222-226). IEEE.

Hao, Y., Li, Y., Dong, X., Fang, L., & Chen, P. (2018). Performance analysis of consensus algorithm in private blockchain. In *2018 IEEE Intelligent Vehicles Symposium (IV)* (pp. 280-285). IEEE.

Luo, X., Cai, W., Wang, Z., Li, X., & Leung, C. V. (2019). A payment channel based hybrid decentralized ethereum token exchange. In *2019 IEEE International Conference on Blockchain and Cryptocurrency (ICBC)* (pp. 48-49). IEEE.

Barbon, A., & Ranaldo, A. (2021). On The Quality Of Cryptocurrency Markets: Centralized Versus Decentralized Exchanges. arXiv preprint arXiv:2112.07386.

Luo, X., Wang, Z., Cai, W., Li, X., & Leung, V. C. (2020). Application and evaluation of payment channel in hybrid decentralized Ethereum token exchange. *Blockchain: Research and Applications*, 1(1-2), 100001.

Lin, L. X. (2019). Deconstructing decentralized exchanges. *Stan. J. Blockchain L. & Pol'y*, *2*, 1.

Victor, F., & Weintraud, A. M. (2021). Detecting and quantifying wash trading on decentralized cryptocurrency exchanges. In *Proceedings of the Web Conference 2021* (pp. 23-32).

Chen, Y., & Bellavitis, C. (2020). Blockchain disruption and decentralized finance: The rise of decentralized business models. *Journal of Business Venturing Insights*, 13, e00151.

Yin, R. K. (1981). The case study as a serious research strategy. *Knowledge*, *3*(1), 97-114.

Yazan, B. (2015). Three approaches to case study methods in education: Yin, Merriam, and Stake. *The qualitative report*, *20*(2), 134-152.

Aberdeen, T. (2013). Yin, RK (2009). Case study research: Design and methods. Thousand Oaks, CA: Sage. *The Canadian Journal of Action Research*, *14*(1), 69-71.

Stake, R. E. (2013). *Multiple case study analysis*. Guilford press.

Pan, S. L., & Scarbrough, H. (1999). Knowledge management in practice: An exploratory case study. *Technology analysis & Strategic management*, *11*(3), 359-374.

Ogawa, R. T., & Malen, B. (1991). Towards rigor in reviews of multivocal literatures: Applying the exploratory case study method. *Review of educational research*, *61*(3), 265-286.

Lam, P. T., & Law, A. O. (2016). Crowdfunding for renewable and sustainable energy projects: An exploratory case study approach. *Renewable and sustainable energy reviews*, *60*, 11-20.

Flavell, H., Harris, C., Price, C., Logan, E., & Peterson, S. (2019). Empowering academics to be adaptive with eLearning technologies: An exploratory case study. *Australasian Journal of Educational Technology*, *35*(1).

Sali, A. H. A., & Marasigan, A. C. (2020). Madrasah Education Program implementation in the Philippines: an exploratory case study. *International Journal of Comparative Education and Development*.

Tian, Y., Minchin, R. E., Chung, K., Woo, J., & Adriaens, P. (2022a). Towards Inclusive and Sustainable Infrastructure Development through Blockchain-enabled Asset Tokenization: An Exploratory Case Study. In *IOP Conference Series: Materials Science and Engineering* (Vol. 1218, No. 1, p. 012040). IOP Publishing.

SolarCoin. (2021). *SolarCoin is a cryptocurrency that incentivizes a solar-powered planet.* https://solarcoin.org/

WePower. (2021). *Platform connecting energy suppliers, corporate buyers and energy producers for easy, direct green energy transactions.* https://wepower.com/.

Cenfura. (2021). *Energy Settlement Token.* https://xcf-token.io/.

Ziyen. (2020). *ZiyenCoin is an Oil & Energy Security Token Offering (STO).* https://www.ziyen.com/ziyencoin_future_currency_of_oil_industry/.

EVZ. (2021). *The Future of Electric Vehicle Charge Platform.* https://www.evzlife.com/index_en.html.

GNL. (2021). *ONE OF THE WORLD'S BEST DIGITAL ASSETS FOR RENEWABLE ENERGY AND RECYCLING*. https://www.greenlifeenergyglobal.com/.

ImpactPPA. (2018). *Delivering access to clean renewable energy from a mobile device. Pre-pay technology unlocks the vast potential of the emerging markets*. https://www.impactppa.com/.

Sun Exchange. (2018). *Solar Cell vs SUNEX Token Purchases*. https://medium.com/@alias_73214/solar-cell-vs-sunex-token-purchases-5bae939b37e0.

NETZ. (2021). *Invest in Pre Defined and Fully Risk Mitigated Green Energy Assets A Real World Asset Backed Crypto Currency*. https://www.netzcoin.io/.

SolarWind. (2021). *SOLARWIND $SLW Energy is the new currency. SolarWind is leading the charge.* https://solarwindtoken.net/.

Restart Energy. (2021). *Restart Energy Democracy – MWAT Token*. https://restartenergy.ro/en/restart-energy-democracy-mwat-token/.

Electrify. (2021). *Introducing Synergy*. https://www.electrify.asia/.

SunContract. (2021). *SunContract: Electricity Marketplace for P2P energy trading*. https://suncontract.org/.

Megatech. (2021). *Megatech MGT Solar | Asset Backed Cryptocurrency | Invest in green solar energy in South Africa and receive dividends from 8 solar plants.* https://mgtsolar.com/.

Enercom. (2021). *Security Token Offering - Enercom AG*. https://www.enercom.ag/security-token-offering/.

Sun Fund. (2019). *Solar Energy – Vertically Integrated*. https://www.startengine.com/sun-fund-dc.

GridPlus. (2021). *The new standard for hardware wallets*. https://gridplus.io/.

WPP. (2021). *WPP ENERGY GmbH – Green Energy Token*. https://wppenergy.io/.

Efforce. (2021). *Energy Efficiency. Reinvented*. https://efforce.io/.

Powerledger. (2021). *Powerledger Energy Projects*. https://www.powerledger.io/.

United Nations. (2014). *Country classification.* https://www.un.org/en/development/desa/policy/wesp/wesp_current/2014wesp_country_classification.pdf

Kshetri, N., & Voas, J. (2018). Blockchain in developing countries. *It Professional*, *20*(2), 11-14.

Wu, K., Wheatley, S., & Sornette, D. (2018). Classification of cryptocurrency coins and tokens by the dynamics of their market capitalizations. *Royal Society open science*, *5*(9), 180381.

Ferrari, V. (2020). The regulation of crypto-assets in the EU–investment and payment tokens under the radar. *Maastricht Journal of European and Comparative Law*, *27*(3), 325-342.

Toderean, L., Antal, C., Antal, M., Mitrea, D., Cioara, T., Anghel, I., & Salomie, I. (2021). A Lockable ERC20 Token for Peer to Peer Energy Trading. *arXiv preprint arXiv:2111.04467*.

Nadler, P., & Guo, Y. (2020). The fair value of a token: How do markets price cryptocurrencies?. *Research in International Business and Finance*, *52*, 101108.

Lipton, A., Sardon, A., Schär, F., & Schüpbach, C. (2020). 11. Stablecoins, Digital Currency, and the Future of Money. In *Building the New Economy*. PubPub.

Prat, J., Danos, V., & Marcassa, S. (2019). Fundamental pricing of utility tokens.

Crosser, N. (2018). Initial coin offerings as investment contracts: Are blockchain utility tokens securities. *U. Kan. L. Rev.*, *67*, 379.

Pazos, J. (2018). Valuation of utility tokens based on the quantity theory of money. *The Journal of the British Blockchain Association*, *1*(2), 4318.

Baranes, E., Hege, U., & Kim, J. H. (2021). Utility Tokens Financing, Investment Incentives, and Regulation. *Investment Incentives, and Regulation*.

Mendelson, M. (2019). From initial coin offerings to security tokens: A US Federal Securities law analysis. *Stan. Tech. L. Rev.*, *22*, 52.

Subramanian, H. (2019). Security tokens: architecture, smart contract applications and illustrations using SAFE. *Managerial Finance*.

Rivero, A. (2018). Distributed Ledger Technology and Token Offering Regulation. *Available at SSRN 3134428*.

Hines, B. (2020). *Digital Finance: Security Tokens and Unlocking the Real Potential of Blockchain*. John Wiley & Sons.

Lambert, T., Liebau, D., & Roosenboom, P. (2021). Security token offerings. *Small Business Economics*, 1-27.

Liu, C., & Wang, H. (2019). Crypto tokens and token offerings: an introduction. *Cryptofinance and mechanisms of exchange*, 125-144.

Rizal Batubara, F., Ubacht, J., & Janssen, M. (2019, June). Unraveling transparency and accountability in blockchain. In *Proceedings of the 20th Annual International Conference on Digital Government Research* (pp. 204-213).

Reyna, A., Martín, C., Chen, J., Soler, E., & Díaz, M. (2018). On blockchain and its integration with IoT. Challenges and opportunities. *Future generation computer systems*, *88*, 173-190.

Schär, F. (2021). Decentralized finance: On blockchain-and smart contract-based financial markets. *FRB of St. Louis Review*.

Tian, Y., Adriaens, P., Minchin, R. E., Chang, C., Lu, Z., & Qi, C. (2020b). Asset Tokenization: A blockchain Solution to Financing Infrastructure in Emerging Markets and Developing Economies. *Institute of Global Finance Working Paper No. Forthcoming.*

Giungato, P., Rana, R., Tarabella, A., & Tricase, C. (2017). Current trends in sustainability of bitcoins and related blockchain technology. *Sustainability*, *9*(12), 2214.

Uzsoki, D., & Guerdat, P. (2019). *Impact Tokens: A Blockchainbased Solution for Impact Investing*. International Institute for Sustainable Development.

Savelyev, A. (2018). Copyright in the blockchain era: Promises and challenges. *Computer law & security review*, *34*(3), 550-561.

Schwerin, S. (2018). Blockchain and privacy protection in the case of the european general data protection regulation (GDPR): a delphi study. *The Journal of the British Blockchain Association*, *1*(1), 3554.

Morkunas, V. J., Paschen, J., & Boon, E. (2019). How blockchain technologies impact your business model. *Business Horizons*, *62*(3), 295-306.

Collomb, A., & Sok, K. (2016). Blockchain/distributed ledger technology (DLT): What impact on the financial sector?. *Digiworld Economic Journal*, (103).

Tian, Y., Minchin, R. E., Petersen, C., Moayed, E., & Adriaens, P. (2022b). Financing Public-Private Partnership Infrastructure Projects through Tokenization-enabled Project Finance on Blockchain. In *IOP Conference Series: Materials Science and Engineering* (Vol. 1218, No. 1, p. 012027). IOP Publishing.